\documentclass{PHYEAUTH}

\usepackage{graphicx}
\usepackage{amsmath}
\usepackage{amssymb}

\begin{document}

\begin{frontmatter}

\title{Resistively Detected NMR in Quantum Hall States:  Investigation of the anomalous lineshape near $\nu=1$}

\author[address1]{C.R.~Dean\thanksref{thank1}},
\author[address1]{B.A.~Piot}, 
\author[address3]{L.N.~Pfeiffer}
\author[address3]{K.W.~West}, and 
\author[address1]{G.~Gervais},
\address[address1]{Department of Physics, McGill University, Montreal, H3A 2T8 Canada}
\address[address3]{Bell Laboratories, Lucent Technology, Murray
Hill, NJ, 07974 USA}

\thanks[thank1]{Corresponding author.E-mail: deanc@physics.mcgill.ca}


\begin{abstract}

A study of the resistively detected nuclear magnetic resonance (RDNMR) lineshape in the vicinity of $\nu=1$ was performed on a high-mobility 2D electron gas formed in GaAs/AlGaAs. In higher Landau
levels, application of an RF field at the nuclear magnetic resonance frequency coincides with an observed minimum in the longitudinal resistance, as predicted by the simple hyperfine interaction picture.  Near $\nu=1$ however, an anomalous dispersive lineshape is observed where a resistance peak follows the usual minimum. In an effort to understand the origin of this anomalous peak we have studied the resonance under various RF and sample conditions. Interestingly, we show that the
lineshape can be completely inverted by simply applying a DC current. We interpret this as
evidence that the minima and maxima in the lineshape originate from two distinct mechanisms.

\end{abstract}


\end{frontmatter}


\section{Introduction}

Resistively detected nuclear magnetic resonance (RDNMR) has emerged as a powerful tool for studying the integer and fractional quantum Hall states in 2D electronic systems \cite{Desrat:PRL:88(25)-2002,Stern:PRB:70-2004,Gervais:PRL:94-2005,Tracy:PRL:98-2007,Zhang:PRL:98-2007}.
This relatively new technique directly exploits the hyperfine interaction
between the 2D electron gas (2DEG) and surrounding nuclear spins, allowing us to probe the many fascinating 2D electronic
features (crystalline spin texture, quantum transport dynamics, composite boson/fermion particle formation) in a
novel way. However, while RDNMR offers the potential to gain new insight into a wide range of interesting many body
electron physics, several features of the measured signal remain unexplained \cite{Desrat:PRL:88(25)-2002,GervaisPRB}.

It was first demonstrated almost 20 years ago that electron/nuclear spin interactions in the Quantum Hall Effect (QHE) regime can be observed in the longitudinal resistance \cite{Dobers:PRL:61(14)-1988,Kronmuller:PRL:82(20)-1999}.  Electrons couple to the nuclei via the contact hyperfine interaction, A\textbf{I}$\cdot$\textbf{S}, such that under the application of an external magnetic field, B, the total electronic Zeeman energy is given by

\begin{equation}
\label{eqn:Zeeman}
     E_{Z} = g^{*}\mu_{B}BS_{z} + A\left<I_{z}\right>S_{z}
\end{equation}
\noindent
where $g^{*}$ is the effective electronic g factor, $S_{z}$ is the electron spin along the field direction, $A$ is the hyerfine coupling constant, and $\left<I_{z}\right>$ is the nuclear spin polarization.  

In the thermally activated regime the longitudinal resistance is given by $R_{xx}\propto e^{-\Delta/2k_{B}T}$, where the energy gap, $\Delta$, depends on the Zeeman energy. We can rewrite the Zeeman energy as $E_{Z}=g^{*}\mu_{B}(B+B_{N})S_{z}$, where $B_{N}=A\left<I_{z}\right>/g^{*}\mu_{B}$ defines the effective nuclear magnetic field seen by the electrons (Overhauser shift).  In this way we see that the longitudinal resistance is determined in part by the hyperfine interaction coupling, and is therefore sensitive to variations in the nuclear field.    

Since the effective g-factor in GaAs is negative ($g^{*}=-0.44$), the nuclear field, $B_{N}$, is \textit{opposite} to the applied field, B, reducing the Zeeman energy.  Applying transverse radiation at resonance destroys the nuclear polarization (\textit{i.e.} destroying B$_{N}$), which therefore increases the Zeeman gap.  At odd filling factor, for example, where $\Delta$ varies directly with the Zeeman gap, this results in a corresponding decrease in the measured R$_{xx}$,and thereby gives a means to resistively detect the nuclear magnetic resonance condition.

While such a minima is measured at nearly all integer and fractional filling factors in GaAs/AlGaAs quantum hall samples, in the vicinity of $\nu=1$ an anomalous ``dispersive'' lineshape is observed, where on resonance the usual resistance minima is followed by a secondary resistance maxima at slightly higher RF frequency\cite{Desrat:PRL:88(25)-2002,Tracy:PRB:73-2006,Kodera:PhysStatSol:3(12)-2006}. In an effort to
understand the origin of this anomalous peak, which remains unexplained by the simple contact hyperfine
interaction picture, we studied the resonance lineshape near $\nu=1$ under various RF and sample conditions.


\section{Experiment}

Resistively detected NMR was performed on a relatively low density ($1.60(1)\times10^{11}$~cm$^{-2}$), high mobility ($\mu\sim17\times10^{6}$~cm$^{2}$V$^{-1}$s$^{-1}$), 2DEG confined to a 40~nm wide modulation-doped GaAs/AlGaAs quantum well.  The sample was cooled in a dillution refrigerator (base temperature $\sim$17~mK) equipped with a 9 Tesla magnet.  Treatment with illumination from a red LED was used during the initial cool down.  Cooling the electrons in the 2DEG was achieved by thermally anchoring the sample leads to the fridge via powder and RC low pass filters. RuO and CMN thermometry, both calibrated by a fixed point device, were used to monitor the fridge and sample temperatures.  Resistance measurements were performed under quasi-dc conditions ($I=10$~nA, $f=13.5$~Hz) using a standard lock-in technique.  

\begin{figure}[h]
	\begin{center}
	\leavevmode
	\includegraphics[width=1\linewidth]{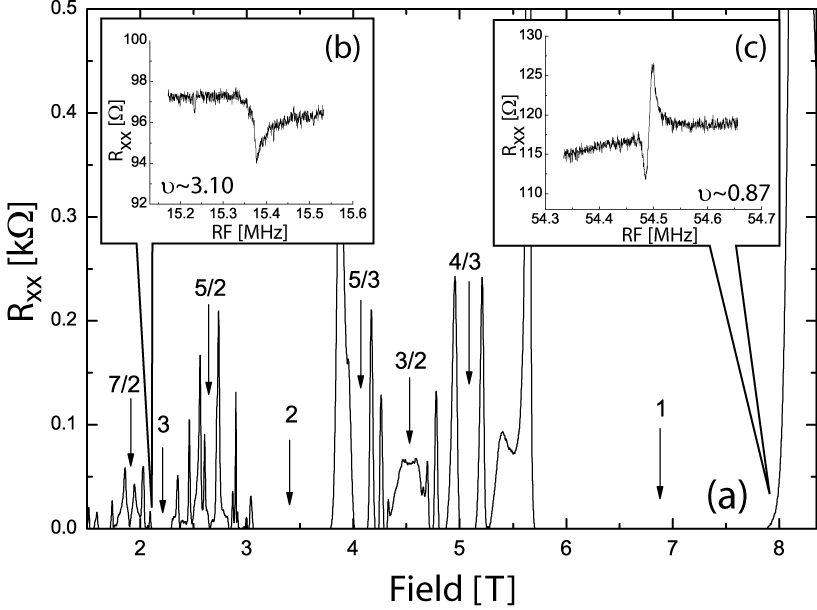}
	\caption{(a) Longitudinal resistance versus magnetic field at T=17~mK.  Typical RDNMR signals are shown inset at filing factors (b) $\nu\sim3.10$, and (c) $\nu\sim0.87$}
	\label{Fig:RxxVsB}
	\end{center}
\end{figure}

All NMR measurements were performed at fixed field with the RF frequency swept through resonance at constant RF power.  RF radiation was applied transverse to the static B field using an 8 turn coil around the sample. The RF frequency sweep rate was 300~Hz/3~sec.  The RF power amplitude is quoted by a decibel scale where 0dBm would correspond to a 440~mV$_{pk-pk}$ sine wave applied to the top of the cryostat.  We typically shine a -13dBm signal which we estimate gives a few ~$\mu T$ field radiated at the sample.  At these RF conditions non resonant electron heating, determined by the corresponding change in the magnetoresistance, was measured to be $\sim$35 mK above the bath temperature (at base).  A magnetotransport measurement at base temperature is shown in Fig. \ref{Fig:RxxVsB}, with typical RDNMR signals shown in the inset.


\section{NMR lineshape near $\nu=1$}

\begin{figure}[h]
	\begin{center}
	\leavevmode
	\includegraphics[width=1\linewidth]{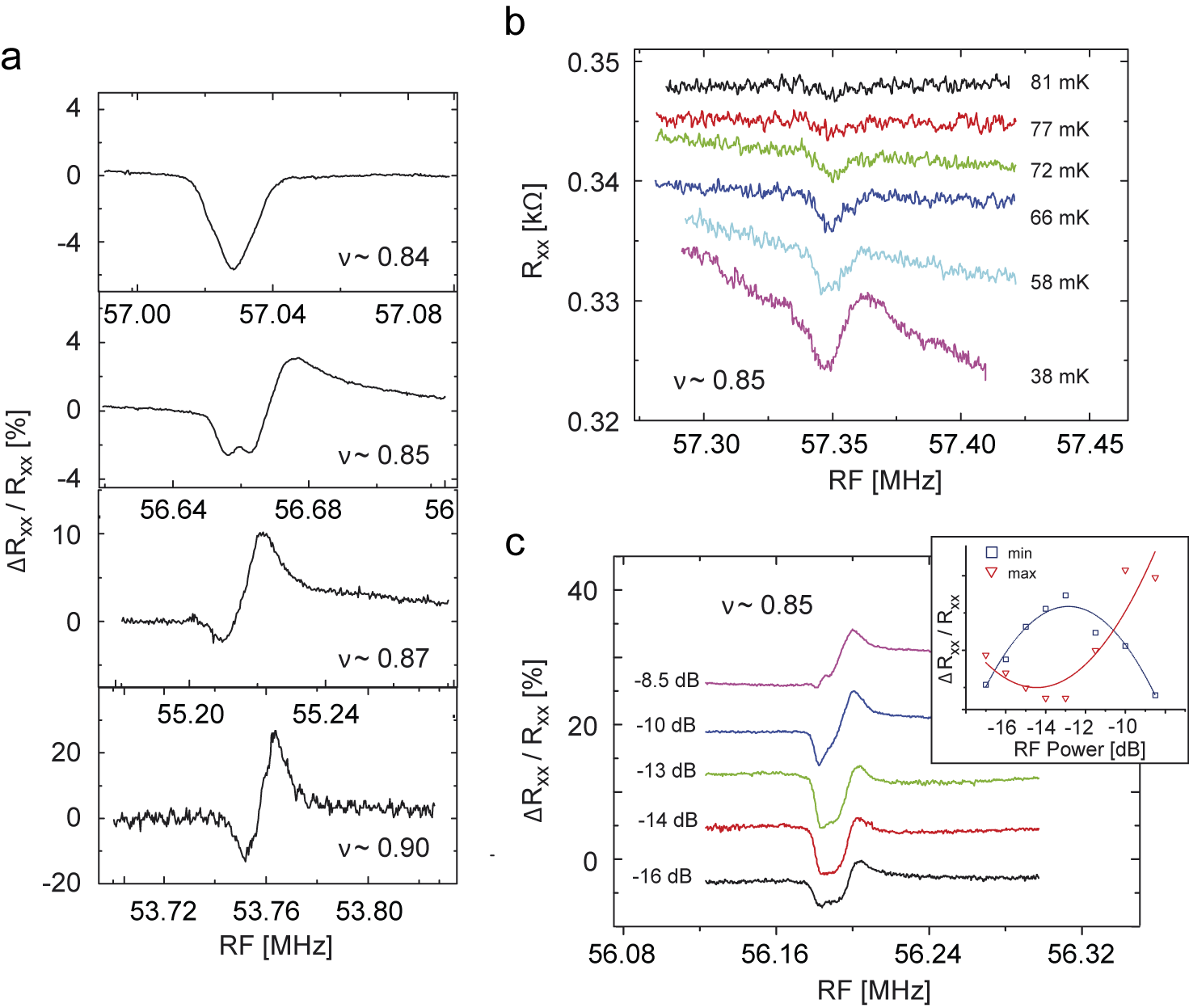}
	\caption{(a)NMR lineshape dependence on (a) filling factor (T$_{e}\sim55~mK$, RF power = -13dBm);  (b) bath temperature ($\nu\sim0.85$, RF power = -17dBm); and (c) applied RF power ($\nu\sim0.85$). (inset shows the evolution of the normalized minima and maxima heights vs power)}
	\label{fig:NMRvsParams}
	\end{center}
\end{figure}

Fig. \ref{fig:NMRvsParams} shows the resistively detected NMR lineshape in the vicinity of $\nu=1$ under various sample and RF conditions.  In all cases, the resonance condition of the $^{75}As$ nuclei was chosen since this is the most abundant isotope in GaAs and thus gives the strongest signature.

Increasing the RF power causes an increase in the electron temperature (seen by a corresponding increase in the measured R$_{xx}$).  However, the lineshape evolution versus RF power (Fig. \ref{fig:NMRvsParams}c) shows different behaviour to that observed when changing the bath temperature at fixed RF power (Fig. \ref{fig:NMRvsParams}b).  In Fig. \ref{fig:NMRvsParams}b, as the bath temperature is increased from 38~mK to 81~mK, the signal goes from a dispersive like shape, to a minima only, before vanishing altogether.  Conversely, as the applied RF power is increased (Fig \ref{fig:NMRvsParams}c) the signal goes from a lineshape in which the minima dominates to a strongly dispersive shape.  Increasing the strength of the transverse RF field radiated on the sample, in the regime where the field is already quite small, can more efficiently destroy the nuclear polarization, causing a larger change in the Zeeman gap at resonance and thus a larger measured response in the resistivity.  However, non-resonant heating caused by application of the RF radiation will naturally diminish nuclear polarization, which follows a Boltzman temperature distribution.  Likely, a competition between these two effects gives rise to the two different trends observed.

Kodera \textit{et al.} recently published a study of the dispersive lineshape as a function of filling factor and sample temperature \cite{Kodera:PhysStatSol:3(12)-2006}.  While the same qualitative observations are reproduced here, the details differ.  Kodera found the lineshape to transition from dispersive to a minima upon moving away from $\nu=1$, in agreement with the trend shown in Fig. \ref{fig:NMRvsParams}a.  However, Kodera shows a dispersive shape persisting all the way to $\nu\sim0.82$ whereas here the signal shows a strong lorentzian shaped minima at $\nu\sim0.84$.  Similarly, Kodera finds the NMR signal diminishing to zero beyond $\sim$200~mK bath temperature, whereas here, at a similar filling factor, the signal dies out at a much lower temperature of $\sim$100~mK. 
Differences between these two studies might simply be sample dependent, as the 2DEG examined here has a much higher mobility. However this might  also be attributable to differences in the amplitude of the transverse RF field.  For example, while we found that increasing the RF power changes the resonance from a minima to dispersive shape, the RF amplitude where this transition occurs varies with filling factor.  We also found that the RDNMR signals persist to higher temperature when applying higher RF power.

\section{Current induced lineshape inversion}

In a recent study, Tracy \textit{et al.} argued the dispersive lineshape could be understood from the perspective of electronic temperature \cite{Tracy:PRB:73-2006}.  They found that as the filling factor was changed through a region where $dR_{xx}/dT$ changes sign, the shape of the signal undergoes an inversion.  Since the lineshape so closely tracks the change of sign in $dR_{xx}/dT$, it was argued the signal results from a cooling of the electron gas on the low frequency side of the resonance, followed by a heating on the high frequency side.  However, the mechanism causing cooling versus heating around the resonance remains unexplained.  

\begin{figure}[h]
	\begin{center}
	\leavevmode
	\includegraphics[width=1\linewidth]{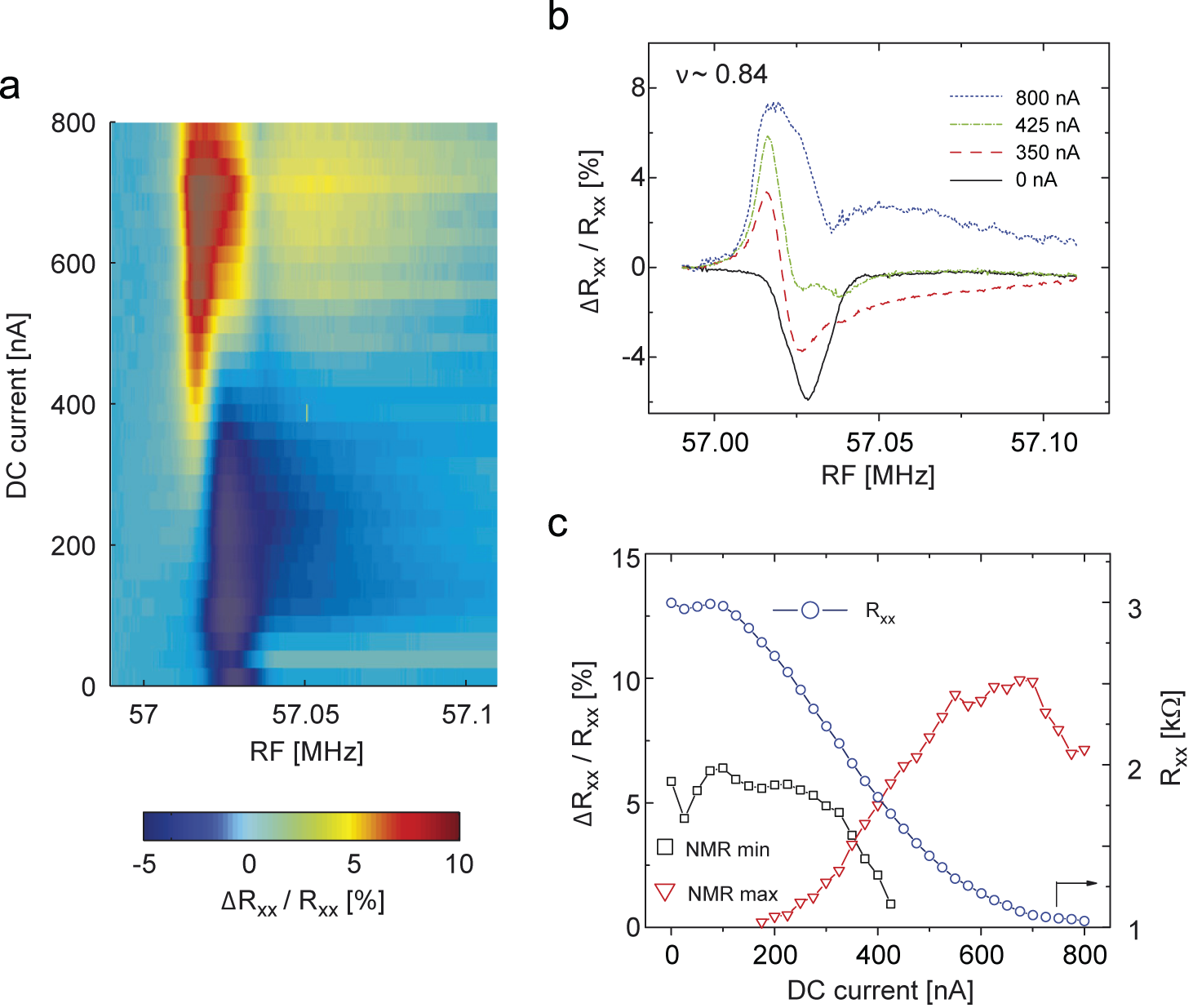}
	\caption{(a)2D contour map showing inversion of the NMR signal with increasing DC current. (b) Evolution of the NMR lineshape at selected DC current values. (c) Squares and triangles indicate the size of the minima and maxima signals with increasing current.  Circles indicate the corresponding off-resonant longitudinal resistance.  All data at $\nu\sim0.84$, and bath temperature of $\sim$27~mK}.
	\label{fig:NMRvsIdc}
	\end{center}
\end{figure}

Fig. \ref{fig:NMRvsIdc} shows the NMR signal as a function of applied DC current at fixed filling factor ($\nu\sim0.84$).  Note that the signal at I$_{dc}=0$~nA is the same ``minima only'' signal shown in Fig. \ref{fig:NMRvsParams}a.  Application of a DC current heats electrons in the 2DEG which results in a narrowing of the of the QHE plateaus.  Sweeping the DC current at a fixed filling factor therefore causes R$_{xx}$ to vary in a correlated way.  This is shown by the circles in Fig. \ref{fig:NMRvsIdc}c, which indicate the off resonance value of R$_{xx}$ as a function of current.  At I$_{dc}=0$ we are already high on the flank of the $\nu=1$ plateau ($\nu\sim0.84$).  As the current is increased, the plateau beings to collapse and the $R_{xx}$ peak shifts to the left (lower magnetic field).  Correspondingly, the measured magnetoresistance first increases to the height of the shifting peak, then begins to fall.  The most striking effect of increasing the current is a complete inversion of the NMR signal. 

Examining the evolution of the inversion, shown in Fig. \ref{fig:NMRvsIdc}b, reveals several interesting features.  Importantly, the position in {\it frequency} of the inverted  maxima {\it does not coincide} with the initial minima, being shifted downwards by $\sim$10~kHz, which is of order of the expected 
Knight shift in this filling factor region. Furthermore, the peak appears to develop alongside the minima, giving a dispersive lineshape over a range of nearly 200~nA DC current (centered around $\sim$400 nA).  It has been well established that applying a DC current to quantum hall samples heats the 2DEG such that for a given applied current, the shape of the QHE magnetoresistance can be reproduced by heating the bath  with negligible current applied \cite{Wei:PRB:50(9)-1994,Chow:PRL:77(6)-1996}.  Therefore, the sign of the current dependance of the magnetoresistance should correlate directly with its temperature dependance.  The \textit{current induced} signal inversion observed here occurs in a transitional region, near $\sim$ 400 nA,  where $dR_{xx}/dI$ (and therefore $dR_{xx}/dT$) is always negative, and thus does not seem to coincide directly with any sign change in $dR_{xx}/dI$ ($dR_{xx}/dT$).  

It should be noted that the data presented here was acquired at constant applied field $H_{0}$ and so the resonance condition is expected to remain unchanged since the NMR frequency is given simply by $f_{NMR}=\gamma H_{o}$ ($\gamma$ is the gyromagnetic ratio of the nuclei).  The sizable shift in the position of the resonance, together with the evolution from a minima only, to dispersive, to maxima only lineshape, suggests this inversion is not a simple temperature correlated flip of the signal, but instead the result of a more complex process, possibly involving two distinct mechanisms. Most likely, the minima can be understood by the hyperfine interacting picture, whereas the shifted maxima appears to be highly specific to the sample condition, and filling factor. Furthermore, despite the significant heating caused by the applied DC current (the effective electronic temperature at I$_{dc}=$ 800~nA was estimated to be $\sim$130~mK) the NMR signal remains surprisingly strong, with the $\Delta$R$_{xx}$/R$_{xx}$ measured at I$_{dc}=$ 800~nA nearly 1.5 times larger than at I$_{dc}=$ 0~nA.  This is taken as further evidence that the maxima arises from a distinct process than the minimum. A study is ongoing in an effort to investigate and deconvolve the effect of the DC current and the resulting electron heating.

\section{Conclusion}

A study of the resistively detected NMR lineshape in the vicinity of $\nu=1$ found that the anomalous ``dispersive'' shape depends in an interrelated way on the filling factor, sample temperature, and on the strength of the applied transverse RF field.  We further show that at fixed filling factor near $\nu=1$, inversion of the RDNMR signal can be achieved by the application of sufficiently strong DC current, without loss in the signal strength.

This work has been supported by the Natural Sciences
and Engineering Research Council of Canada (NSERC),
the Canada Fund for Innovation (CFI) and the Canadian Institute for
Advanced Research (CIFAR). 
One of the authors (G.G.) acknowledges receipt
of support from the Alfred P. Sloan Foundation under their fellowship
program.

\end{document}